# A Brief Overview of the UML Scientific Profile

Vasyliy I. Gurianov, e-mail: vg2007sns@rambler.ru

**Abstract.** *Model-Driven Engineering is of great importance not only for information systems and software development but also for building computer models and simulation. Model-oriented engineering deals with identifying, analyzing, and describing core concepts and limits real-domain constraints through the use of a modeling language that is founded on a set of basic modeling concepts. In this article, we introduce the UML SP object-oriented simulation language and show how to use it to create simulation models as the basis for model-driven simulation engineering.*

## Introduction

Numerical simulation is widely used in scientific research. It is easy to see that numerical simulation is not the same as simple simulation, which is used in tools such as AnyLogic or ARENA. Why is simulation modeling not used in the scientific field? In our opinion, the reason is that there is no clear understanding of how to build simulation models in the scientific field. To create scientific models, a well-founded and logically rigorous simulation methodology is required. Unified Process (UP) could be such a methodology for Modeling and Simulation (M&S).

In this paper we propose implementing UML-profile, called UML SP (UML Scientific Profile) and originally proposed in [5, 6], to carry out simulation modelling. This profile would allow for the use of the Unified Process to create simulation models. The Unified Process is a tried-and-tested software development methodology. In this article, we give a brief overview of the UML SP. An extended version of this paper is available as a live document [7].

This paper is organized as follows: Section 1 contains a brief description of the UML SP language. Section 2 demonstrates a simple example simulation model in UML SP. Finally, Section 3 presents the paper's conclusions.

## 1. Syntactics of UML SP

Model-Driven Engineering (MDE), also called model-driven development, is a well-established paradigm in IS&SE (Information Systems and Software Engineering) from the Object Management Group [9]. It is natural to apply the ideas of MDE to simulation engineering as well, see, e.g., the overview presented in Cetinkaya and Verbraeck [1].

Due to their great expressivity and their wide adoption as the standard for modelling, UML Diagrams now appear to be the best choices for simulation modelling. However, since they have not been specifically designed for this purpose, we may have to restrict, modify and extend them in an acceptable manner. This issue is discussed below in further detail.

The UML SP has been used for the development of simulation models. It is a profile of UML. The UML SP profile contains a set of stereotyped classes that support the design of conceptual models according to the decomposition principle. Moreover, the profile also contains some constraints restricting the way the modelling elements may be related. UML SP models do not depend on programming languages.

UML-diagrams have a double semantic in UML SP. In the computational semantic, the class diagram serves as a model of the program. In the problem domain semantic, the class diagram is the ontology. Class is considered as a frame as per Marvin Minsky (more precisely, as in the Protégé system), where the name of the class is the name of the frame. The tagged value "Concept" defines the notion from the problem domain. Class Attributes are slots of the frame. Operations are procedures of the frame. Attributes and Operations also define Concepts.

UML SP allows for the use of UP to develop simulation models. The methodology is applicable for OOS (Object-Oriented Simulation) and ABS (Agent-Based Simulation). This UP modification is called SSP (Simulation with Scientific Profile). SSP includes several models: Use-Case, Analysis, and Design.

The main artefact is the Analysis model. The base method of UML SP is the decomposition principle, as in the IDEF0 (or SADT) methodology. The syntax of UML SP consists of fifteen stereotypes for the execution of decomposition. We use the UML extension mechanism of a UML profile for defining a simulation modelling language whose elements represent the decomposition artefacts.

We view an Analysis model as a conceptual MDE model. There is overall agreement that conceptual modelling is an important first step in a simulation engineering project [11]. However a discussion [13] revealed that there is no agreed-upon definition of what is a conceptual model in M&S. We agree with the following definition [12]: "a formal specification of conceptualization", or "an ontological representation of the simulation that implements it". This corresponds to what is called a domain model in MDE. This definition comes closest to the view taken in this paper.

A similar approach was proposed in the works [2, 3]. The authors presented a Unified Foundational Ontology (UFO) based on theories from Formal Ontology, Cognitive Psychology, Linguistics, Philosophy of Language, and Philosophical Logic. If the set of UML SP stereotypes is created on the decomposition principle then a set of ontoUML stereotypes is selected based on the UFO. In terms of the level of abstraction, UML SP occupies an intermediate position between ontoUML and SysML [10].

## 2. Simulation model example

We'll look at an example to illustrate the use of the profile. Consider the typical service queue from [4]. This will allow you to see the similarities and differences in the use of the UML SP and ontoUML profiles.

In the example service queue system, customers arrive at the service desk at random times, where they have to wait in line when the service desk is busy. Otherwise, when the support service is not busy, it is immediately served by a clerk. Each time the service is completed, the next request from the queue, if any, will be served.

A starting conceptual model of this system may appear as shown in Figure 1.

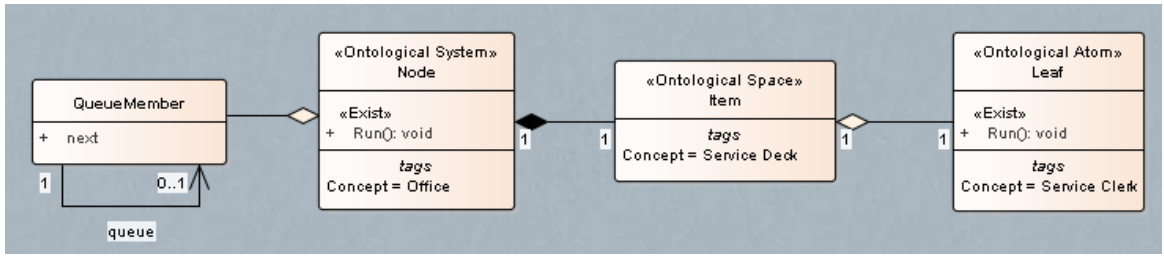

Fig. 1: Conceptual classes for the simulation model.

The SSP modeling guidelines require that we applied a *Composite* pattern (by GoF). In UML SP, the conceptual model is the ontology. The ontology of the Queue System is depicted in Figure 2 (we use the C++ syntactic). In this brief overview, we will not explain the purpose of stereotypes, since their meaning is intuitively clear from the diagram. For a formal definition see [6, 7].

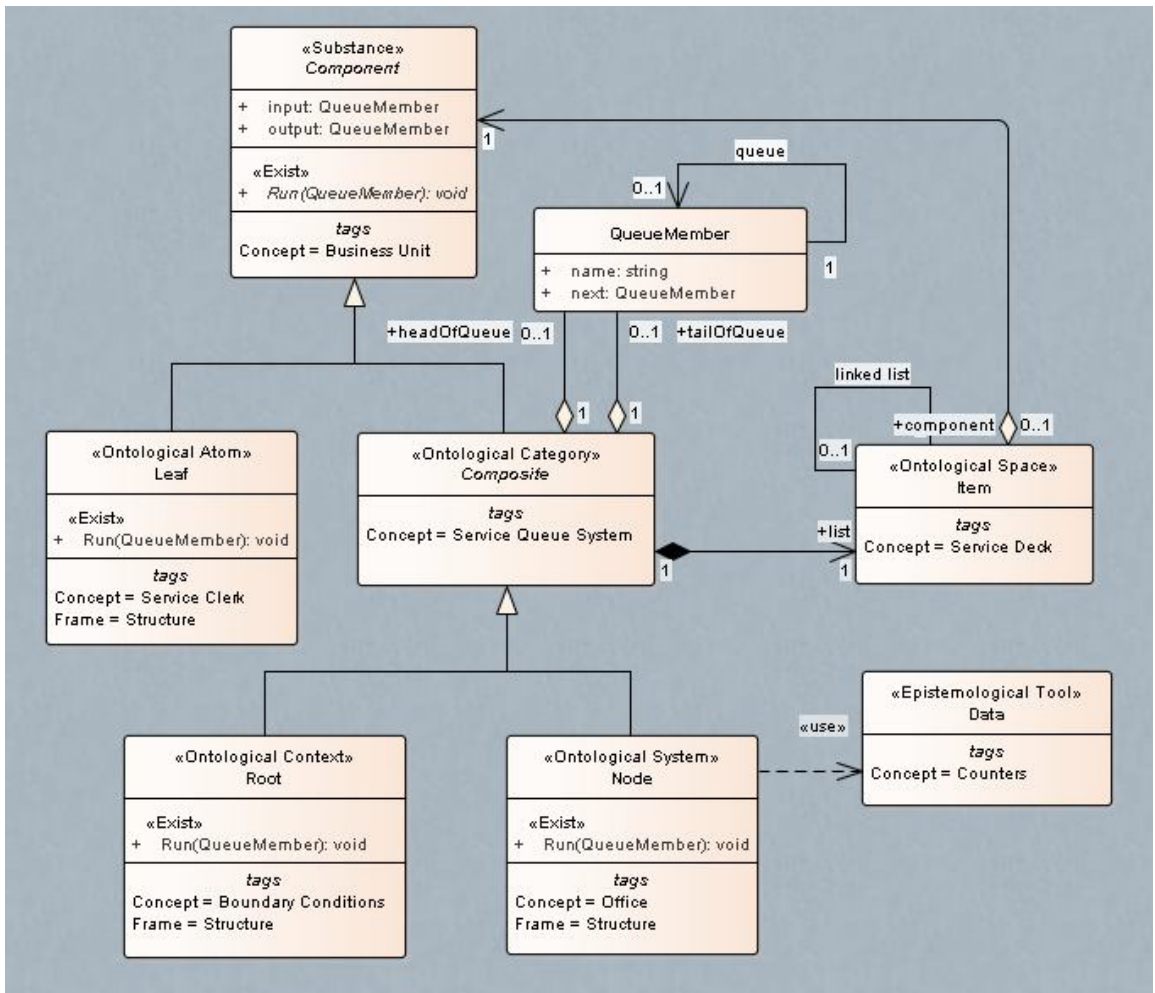

Fig. 2: Ontology for the simulation model.

Computational semantics: the class diagram is a variant of a *Composite* pattern.

In problem domain semantics, the ontology defines the following modelling concepts.

A 'Component' frame defines the 'Business Unit' concept. It has 'input' and 'output' attributes, i.e. it is an open system.

A 'Leaf' frame define 'Service Clerk' concept.

A 'Composite' frame defines the 'Service Queue System' concept and has a 'list' slot. It is an abstract class. The 'list' slot defines the 'Office Space' concept. Both slots 'head' and 'tail' define the 'Queue' concept. Class 'QueueMember' is not a frame. It is included to define the 'Queue'. This class has 'name' and 'next' attributes.

A 'Root' frame defines the 'Boundary and Initial Conditions' concept. It is a boundary for the system.

A 'Node' frame defines the 'Office' concept and it is a concrete class.

Both frames 'Leaf' and 'Root' are symmetric. If we select the clerk as an observer then the 'Root' frame take the atomic frame.

Class operations define rules for changing slots and also define concepts. A critical meaning is given to the operations of the «Exist» stereotype. These operations determine the course of model time.

As a rule, ontology alone is not enough to describe a model. For example, it is required to define the operations algorithm in some detail. One may use communication diagrams for this purpose, Figure 3 .

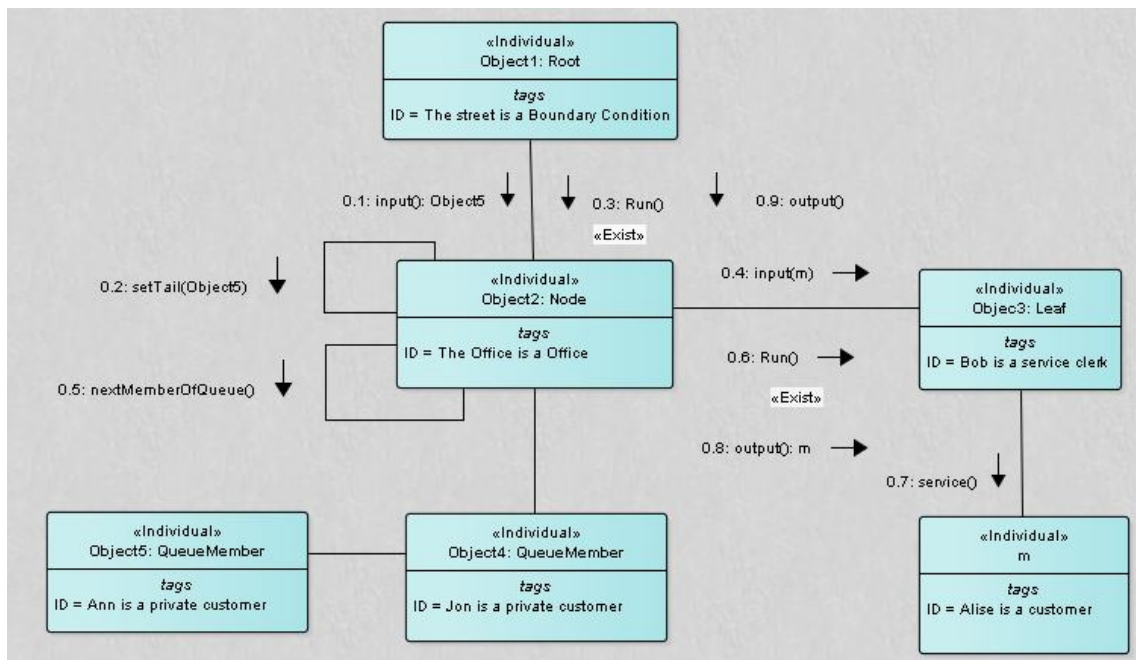

Fig. 3: The messaging order.

Figure 3 describes one tick of the simulation (single call to operation «Exist» of the 'Node' class). The initial state is of two private customers in the queue. A private customer enters the office from the street. The customer stands at the end of the queue. A customer approaches the clerk. The queue is shifted. The clerk is serving the customer. The customer leaves the clerk and the office. According to the definition of the «Exist» stereotype, events are not ordered in time, but what is important is the order in which objects exchange messages.

# Conclusion

To sum up after everything has been detailed. The core concept of UML SP is a communicative paradigm: any object of reality can be described as a system of elementary communicative acts.

The basic idea of the usage of UML SP is a description of the object of study as a semantic net. The semantic net is the non-numeric model. UML SP models may be used for modelling objects with absent mathematical models.

UML SP allows the Unified Process to be used in simulation engineering. This creates a solid foundation for scientific simulation. The current version of the language is UML2 SP ver. 1.1.

The proposed approach is interesting because it allows one to describe the object of study as a semantic network and we can do so without a mathematical model. Thus, the semantic networks offer an alternative to mathematical modelling. This is especially valuable when the mathematical model is unknown. Note that the transformation of a mathematical model into a semantic network is, in general terms, far from a trivial operation. Such is also true for the inverse transformation. Semantic networks are currently used to represent knowledge in AI (Artificial Intelligence). However, as we have shown, semantic networks may be used to create scientific models as well. In our opinion, this line of research deserves special attention.

A real-world example of using the UML SP for scientific simulation can be found in our article [8].